\documentclass{article}

\usepackage{arxiv}

\usepackage[T1]{fontenc}    
\usepackage{hyperref}       
\usepackage{url}            
\usepackage{booktabs}       
\usepackage{amsfonts}       
\usepackage{nicefrac}       
\usepackage{microtype}      
\usepackage{lipsum}
\usepackage{siunitx}

\usepackage{graphicx}
\usepackage{subfig}
\usepackage{tabularx}
\usepackage{slashbox}

\title{Noncontact Thermal and Vibrotactile Display Using Focused Airborne Ultrasound}

\author{
  Takaaki Kamigaki\thanks{\url{https://hapislab.org/takaaki_kamigaki}} \\
  Graduate School of Frontier Sciences\\
  The University of Tokyo\\
  5-1-5 Kashiwanoha, Kashiwa-shi, Chiba-ken, Japan \\
  \texttt{kamigaki@hapis.k.u-tokyo.ac.jp} \\
   \And
  Shun Suzuki \\
  Graduate School of Frontier Sciences\\
  The University of Tokyo\\
  5-1-5 Kashiwanoha, Kashiwa-shi, Chiba-ken, Japan \\
  \texttt{suzuki@hapis.k.u-tokyo.ac.jp} \\
  \And
  Hiroyuki Shinoda  \\
  Graduate School of Frontier Sciences\\
  The University of Tokyo\\
  5-1-5 Kashiwanoha, Kashiwa-shi, Chiba-ken, Japan \\
  \texttt{hiroyuki\_shinoda@k.u-tokyo.ac.jp} \\
}

\begin{document}
\maketitle

\begin{abstract}
In a typical mid-air haptics system, focused airborne ultrasound provides vibrotactile sensations to localized areas on a bare skin. Herein, a method for displaying thermal sensations to hands where mesh fabric gloves are worn is proposed. The gloves employed in this study are commercially available mesh fabric gloves with sound absorption characteristics, such as cotton work gloves without any additional devices such as Peltier elements. The method proposed in this study can also provide vibrotactile sensations by changing the ultrasonic irradiation pattern. In this paper, we report basic experimental investigations on the proposed method. By performing thermal measurements, we evaluate the local heat generation on the surfaces of both the glove and the skin by focused airborne ultrasound irradiation. In addition, we performed perceptual experiments, thereby confirming that the proposed method produced both thermal and vibrotactile sensations. Furthermore, these sensations were selectively provided to a certain extent by changing the ultrasonic irradiation pattern. These results validate the effectiveness of our method and its feasibility in mid-air haptics applications.
\end{abstract}
\section{Introduction}
Recently, mid-air haptics technologies have attracted substantial interest because they can produce tactile sensations without any physical contact or the need for wearing any devices. Airborne ultrasound phased arrays (AUPAs) are one of the most practical devices in mid-air haptics. They can produce vibrotactile sensations on bare skin based on acoustic radiation pressure~\cite{iwamoto2008non, hoshi2010noncontact}. The stimulus area, a focal point generated by AUPAs, can be down to wavelength, i.e., approximately $\SI{8.5}{mm}$ at $\SI{40}{kHz}$ (for typical AUPAs). It can be generated at an arbitrary position and controlled electronically. Complex stimulus patterns, such as the shapes of various objects based on multi-foci~\cite{carter2013ultrahaptics} and focal points with time-division~\cite{korres2016haptogram}, can be produced via the proper drive control of the transducers in AUPAs.
\par
Most existing studies that employ AUPAs have developed applications that can only produce vibrotactile sensations. The realization of other types of haptic sensations using AUPAs, in addition to vibrotactile sensations, can expand the range of applications of AUPAs and contribute to the evolution of mid-air haptics technologies. In this paper, we propose a method to produce thermal sensations using AUPAs, in addition to producing vibrotactile sensations. 
\par
The proposed method requires users to wear gloves to produce thermal sensations, whereas existing methods in mid-air haptics do not have such requirements. This is a limitation of the proposed method; however, the gloves that are required in the proposed method are ordinary ones, such as cotton work gloves that absorb ultrasound, without any additional devices such as Peltier elements. Additionally, vibrotactile sensations can be produced by changing ultrasonic irradiation patterns. We can find some practical applications where wearing gloves are acceptable. For example, at numerous factories, workers wear cotton work gloves while they work. Our method can be employed to prevent the inadvertent intrusion of workers' hands into dangerous zones by imparting a hot sensation as a danger alert. Our method can also be applied to surgery support by improving the surfaces of surgical gloves to absorb sound, where the glove remains disposable and battery-less.
\par
Thermal displays in mid-air haptics, such as methods that employ infrared lasers~\cite{meyer1976laser} and thermal radiation~\cite{saga2015heathapt}, have been proposed. A method that uses lasers can also display a tactile sensation similar to a mechanical tap when an elastic medium is attached to the skin~\cite{lee2015mid}. A generic comparison of the proposed method with the aforementioned laser-based methods is not straightforward; however, it is certain that the proposed method is the easiest, with no additional cost, with regard to application in a scenario where a worker wearing cotton gloves is already being aided by ultrasound mid-air haptics. A method for providing a cold sensation using AUPAs has been proposed; however, providing warm and hot sensations was beyond the scope of this method~\cite{nakajima2018remotely}.
\par
Herein, we report basic experimental investigations regarding the proposed method. We conducted two kinds of experiments: the first involves temperature measurements on the glove and the skin surface when the glove was exposed to ultrasound. The second is a perceptual experiment for confirming that our method provides thermal and vibrotactile sensations and that it can display these sensations selectively by changing the ultrasonic irradiation patterns.
\section{Proposed method}
\begin{figure}[b]
  \centering
  \includegraphics[width = 6cm]{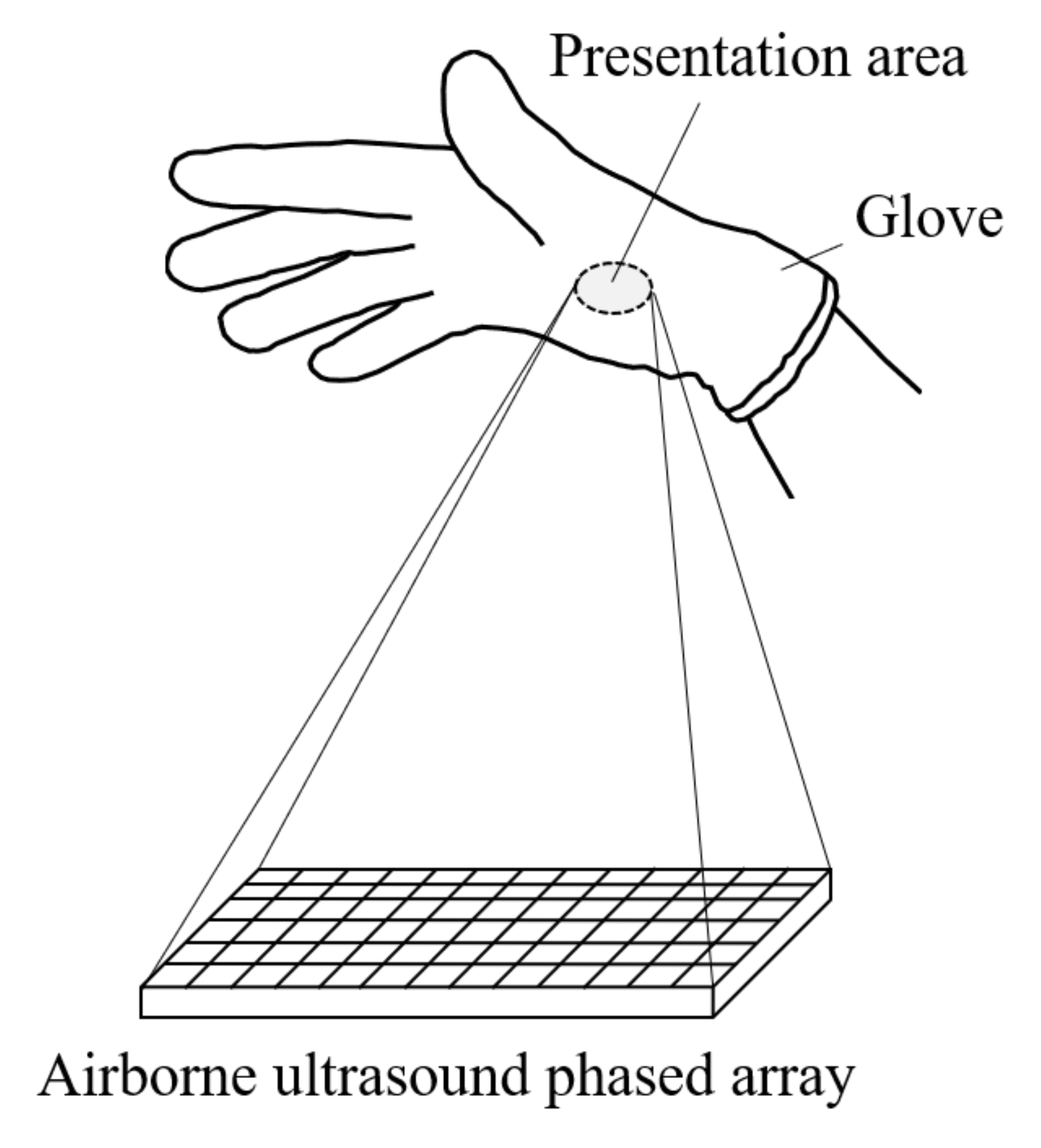}
  \caption{Illustration of the proposed method}
 \label{ProposedMethod}
\end{figure}
Figure~\ref{ProposedMethod} shows an illustration of the proposed method. Both thermal and vibrotactile sensations are produced at a focal point by AUPAs on a glove. Any type of glove can be used in this method, as long as it possesses sound absorption characteristics. A cotton work glove is used in this study. The production of thermal sensations in this method is based on the phenomenon of sound absorption: heat is generated in the focal point on the glove owing to the absorption of sound, which in turn supplies heat to the skin at the contact points of the skin with the glove via the heat conduction. In addition, the generation of vibrotactile sensations utilizes the acoustic radiation pressure of the ultrasound on the glove. 
\par
These sensations can be provided selectively by two modes of irradiation: static pressure (SP) mode where constant-amplitude ultrasound is irradiated and amplitude modulation (AM) mode where the ultrasound is modulated at $\SI{150}{Hz}$. The acoustic absorption coefficient of the glove and the acoustic power at the focal point determine the temperature of the glove exposed to ultrasound. In SP mode, the ultrasound generates heat on the glove while inducing no vibrotactile sensations. In AM mode, the modulated ultrasound produces vibrotactile sensations. The modulation frequency of $\SI{150}{Hz}$ is selected so that the mechanoreceptors are excited efficiently~\cite{hasegawa2018aerial}. The glove temperature also rises in AM mode. However, it is possible to adjust the amplitude and duration of the ultrasound to produce only vibrotactile sensations without thermal sensation. The irradiation duration and amplitude should be adjusted to generate only vibrotactile sensations.
\par
We confirmed the feasibility of the selective stimulation in the following experiments. 
\section{Experiments}
\subsection{Heat generation on glove surface and skin surface}
\begin{figure*}[!b]
\centerline{
\subfloat[]{\includegraphics[height=5.5cm]{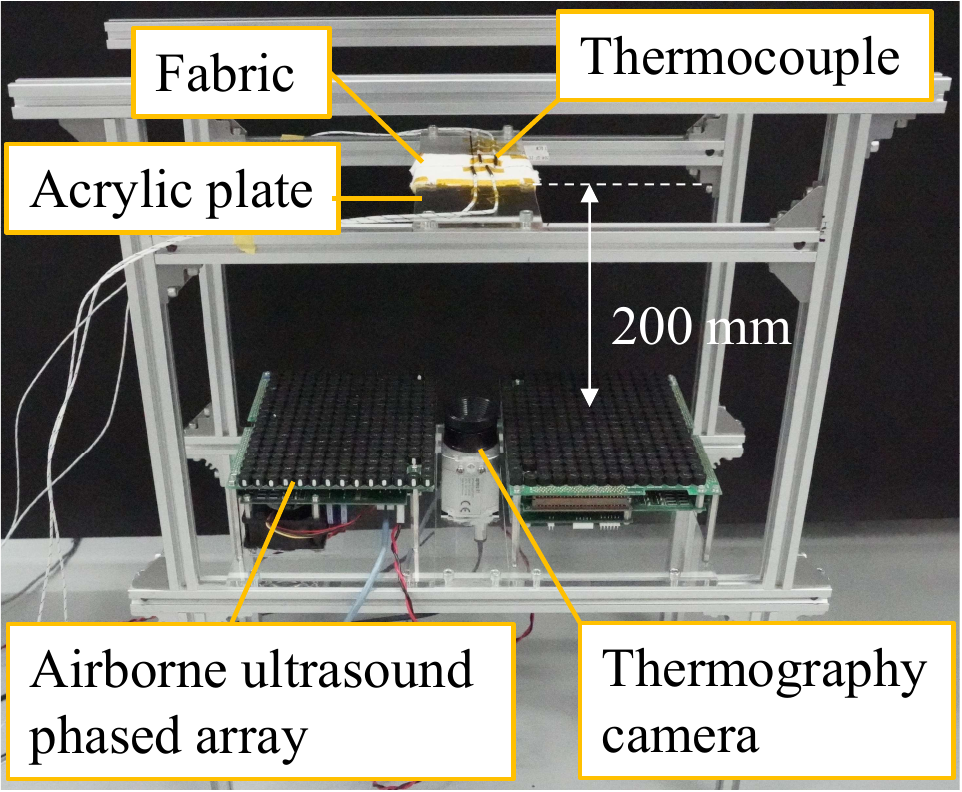}
}
\hfil
\subfloat[]{\includegraphics[height= 5.5cm]{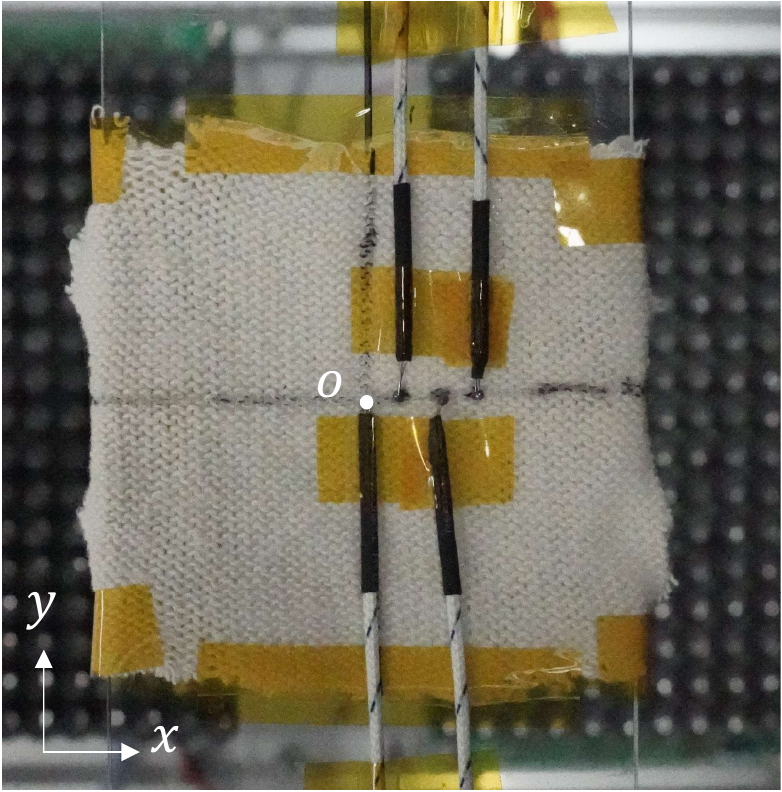}
}
}
\caption{Photograph of the experimental setup. (a) Complete view of the experimental setup, (b) enlarged view of thermocouples and fabric cut from a cotton work glove.}
\label{ExperimentalSetup}
\end{figure*}
First, we measured the temperature elevation of a cotton work glove that is in contact with a human palm. The glove was exposed to ultrasound, and the temperature was measured both on the surface of the glove and the surface of the palm. 
\par
Figure~\ref{ExperimentalSetup} shows the experimental setup. We placed a fabric that was cut from a cotton work glove on an acrylic plate with square perforations of $\SI{50}{mm}$ side lengths. Subsequently, a hand was placed on the fabric to simulate the scenario of a user wearing the glove. We operated two $\SI{40}{kHz}$ AUPAs comprising $498$ transducers (TA4010A1, NIPPON CERAMIC CO., LTD.) such that a focal point was generated on the fabric surface $\SI{200}{mm}$ above the surfaces of the AUPAs. The center of the focal point was considered as the origin point, $(x, y) = (0, 0)\,\SI{}{mm}$, for the measurement. A thermography camera (OPTPI450O29T900, Optris) was employed to measure the temperature distribution on the surface of the fabric irradiated with ultrasound. Simultaneously, four thermocouples discretely arranged between the palm and fabric measured the temperature elevation on the skin. A temperature logger (SHTDL4-HiSpeed, Ymatic Inc.) connected to the thermocouples collected the data with a sampling period of $\SI{10}{ms}$. The thermocouples were placed at 5\,mm intervals in the x-direction, i.e., at $(x, y) = (0, 0), (5, 0), (10, 0)$, and $(15, 0)\,\SI{}{mm}$. The ultrasonic exposure patterns were SP mode and AM mode at $\SI{150}{Hz}$, and the exposure time was restricted to $\SI{10}{s}$ in order not to induce pain and burn injuries.
\par
\begin{figure*}[!t]
\centering
\subfloat[]{\includegraphics[clip, width=14cm]{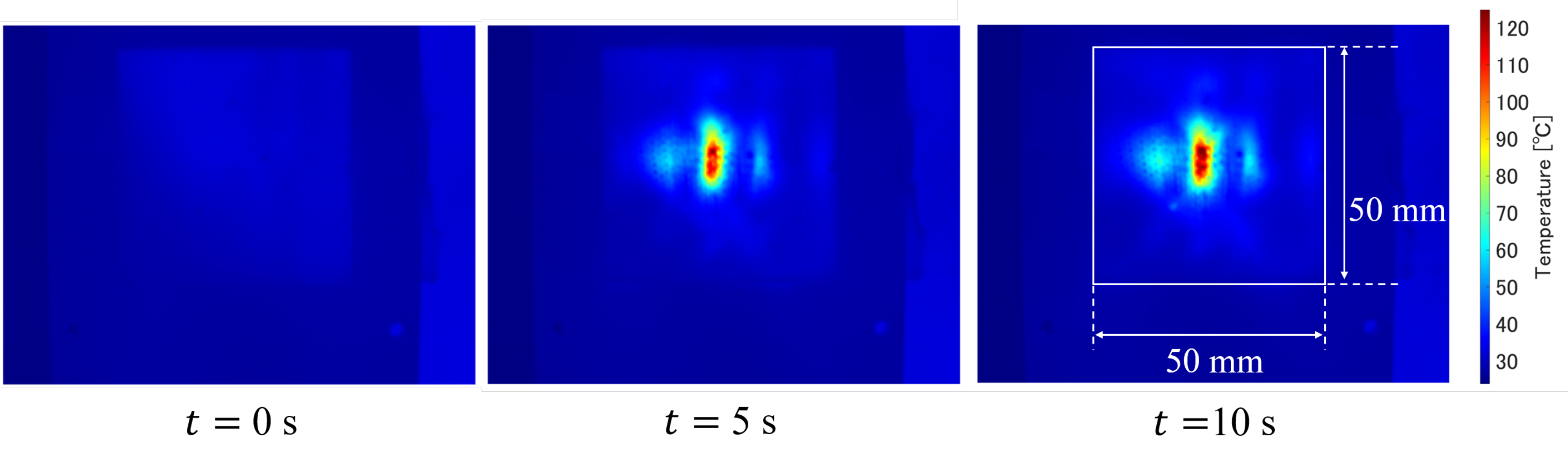}
}
\\
\subfloat[]{\includegraphics[clip, width=14cm]{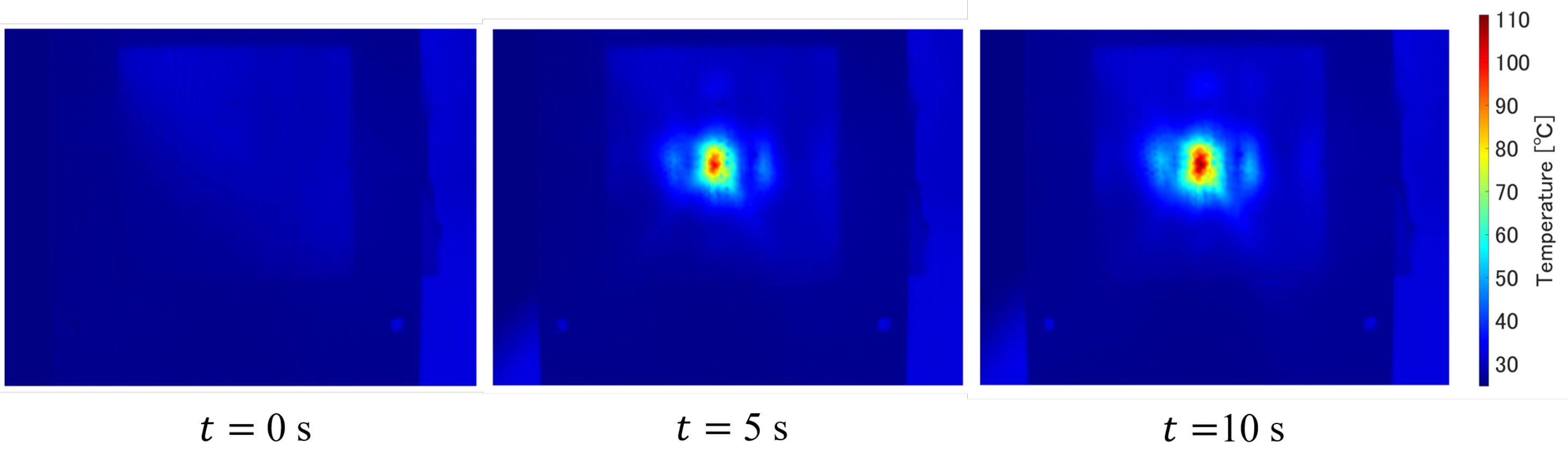}
}
\caption{ Thermal images of observed area in the cases of (a) SP mode, (b) AM mode at $\SI{150}{Hz}$}
\label{MeasuredGloveSurface}
\end{figure*}

Figure~\ref{MeasuredGloveSurface} shows the images of the surface of the fabric that was exposed to the ultrasound over time, from which it can be clearly observed that the proposed method heated the localized area. The variation in the sizes of the focal points obtained for SP and AM modes can be attributed to distance deviation between the fabric and AUPA surfaces caused by the pressing force variation. Although some sidelobes were formed, the obtained temperature distribution is reasonable because the sound distribution on the surface with the focal point is formed in accordance with the sinc function~\cite{hoshi2010noncontact}. The measured temperature was higher in the case of SP mode than that in the case of AM mode because the effective acoustic energy of SP mode is double of AM mode.
\begin{figure}[ht]
  \centering
  \includegraphics[width = 12cm]{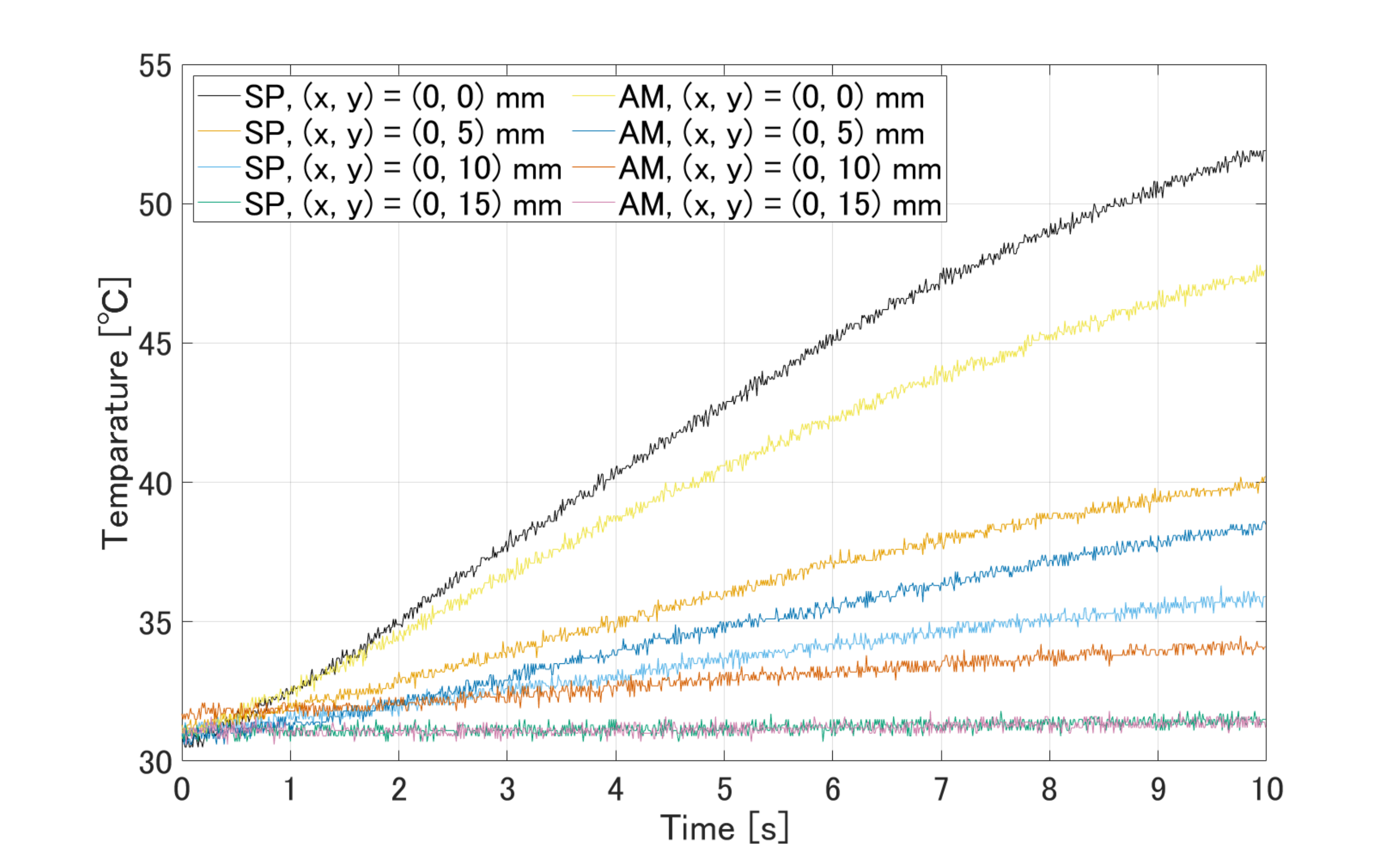}
  \caption{Skin surface temperature measured by each thermocouple}
 \label{MeasuredLogger}
\end{figure}
\par
Figure~\ref{MeasuredLogger} shows the measured results with regard to temperature changes on the skin. These results were obtained by matching the initial average temperatures for the two ultrasonic radiation patterns because the initial average temperatures differed by $\SI{1}{\celsius}$ for the two patterns (SP: $\SI{31.125}{\celsius}$, AM: $\SI{30.125}{\celsius}$). The temperature elevation was highest at the center of the focal point $(x, y)=(0, 0)\,\SI{}{mm}$, and it decreased with an increase in the distance from the center of the focal point, for both radiation patterns. The temperature at $(x, y) = (0, 15)\,\SI{}{mm}$ did not increase for both radiation patterns. Thus, the proposed method heated the surface of the glove locally. Through comparison, it can be seen that the temperature in the case of SP mode was higher than that in the case of AM mode, at every measurement point. These characteristics were inconsistent with the result in Fig.~\ref{MeasuredGloveSurface}. A temperature of $\SI{45}{\celsius}$, which induces the sensation of pain~\cite{jones2008warm} was achieved within $\SI{5.88}{s}$ at the earliest, at $(x, y) = (0, 0)\,\SI{}{mm}$ in the case of SP. Thus, the exposure time must be less than that in this setup. Increasing the number of AUPAs can be considered for achieving a faster increase in temperature because as the input acoustic energy increases, the time needed to induce pain will decrease. Although the relationship between the number of AUPAs and temperature elevation needs to be investigated, it is not considered in this study.
\subsection{Distinguishing vibrotactile and thermal sensations}
Next, we conducted perceptual experiments to confirm whether or not our method can generate both thermal and vibrotactile sensations, as well as that can switch the generated sensation by the SP-AM mode alternation.
\par
\begin{figure}[h]
  \centering
  \includegraphics[height = 6cm]{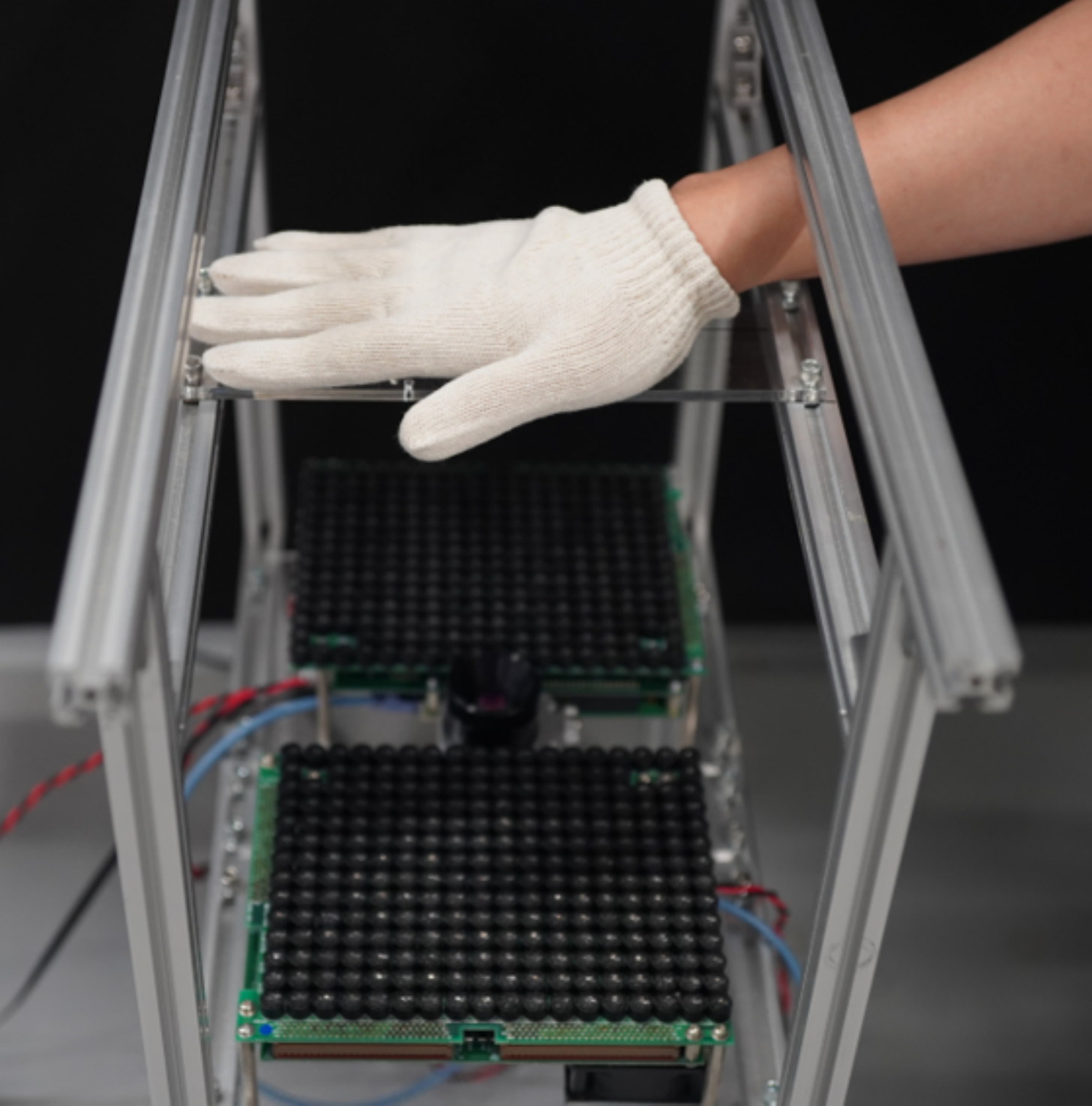}
  \caption{Photograph of perceptual experiment}
 \label{PerceptualExperimentalSetup}
\end{figure}

Figure~\ref{PerceptualExperimentalSetup} shows a photograph of the setup of the perceptual experiments. The experimental setup was the same as that shown in Fig.~\ref{ExperimentalSetup}, except for the thermocouples and the temperature logger. Participants placed their hand, equipped with a cotton work glove, on the acrylic plate. We displayed three patterns: SP mode, AM mode ($\SI{150}{Hz}$), and no irradiation. Each pattern was displayed $10$ times, and the total number of displays was $30$ times. The order of displays was random. In each trial, one of the three patterns were presented for $\SI{5}{s}$. After the presentation time, the participant chose one out of the following four options: ``I only felt heat,'' ``I only felt vibration,'' ``I felt both heat and vibration,'' ``I did not feel anything,'' as compared to what they felt at the beginning of the presentation time. The presentation time of $\SI{5}{s}$ was determined based on the result in Fig.~\ref{MeasuredLogger}, to ensure that the sensation of pain was not induced. Furthermore, the participants waited for $\SI{10}{s}$ after answering before the next trial was started. The participants were made to hear white noise during the experiment to ensure that the sound from the AUPAs did not affect the experimental results. The participants were $9$ men and $1$ woman, with ages ranging from $23$ to $29$ years. The average age of the participants was $25.4$ years.
\par
Table~\ref{Table_PerceptualResults} shows the average results obtained for all participants. These results demonstrate that SP mode caused thermal sensations at a rate of $\SI{98}{\percent}$. Furthermore, AM mode caused vibrotactile sensations at a rate of $\SI{97}{\percent}$, although this was accompanied by thermal sensations at a rate of $\SI{61}{\percent}$. This could be attributed to the deviation of the initial temperature in each trial owing to residual heat from the previous trial. Thus, the proposed method has the possibility of providing only a vibrotactile sensation by adequate AM irradiation time.
\par
In conclusion, the proposed method generated both thermal and vibrotactile sensations. Additionally, these sensations were selectively generated to a certain extent by using different ultrasonic radiation patterns.
\begin{table}[h]
\caption{Average results for 10 participants}
\label{Table_PerceptualResults}
\centering
\begin{tabular}{|l|*{4}{c|}}\hline
\backslashbox{Irradiation patterns}{Answer}
& Heat only & Vibration only & Heat \& Vibration & None\\
\hline
SP mode & $\SI{98}{\percent}$ & $\SI{0}{\percent}$ & $\SI{1}{\percent}$ & $\SI{1}{\percent}$\\\hline
AM mode & $\SI{3}{\percent}$ & $\SI{36}{\percent}$& $\SI{61}{\percent}$ & $\SI{0}{\percent}$\\\hline
No irradiation & $\SI{0}{\percent}$ & $\SI{0}{\percent}$ & $\SI{0}{\percent}$ & $100$\,\%\\\hline
\end{tabular}
\end{table}
\section{Conclusion}
In this paper, we proposed a noncontact method for generating thermal and vibrotactile sensations using focused airborne ultrasound. The proposed method provides thermal sensations to a localized area by irradiating airborne ultrasound to the surface of a hand wearing a mesh glove having sound absorption characteristics. This method also provides vibrotactile sensations by changing the ultrasonic irradiation pattern. It was confirmed that the proposed method locally provided both thermal and vibrotacttile sensations. In addition, our method generated these sensations selectively to a certain extent by varying the ultrasonic radiation pattern. 
\bibliographystyle{ieeetr}
\bibliography{references}
\end{document}